\documentclass[twocolumn,showpacs,prd]{revtex4}
\usepackage{epsf}
\usepackage{dcolumn}
\usepackage{bm}

\usepackage[colorlinks=true,breaklinks=true]{hyperref}
\usepackage[utf8]{inputenc}
\hypersetup{urlcolor=Blue,
	    citecolor=Blue,
	    linkcolor=Blue}
\usepackage[usenames,dvipsnames]{xcolor}        

\newcommand{\GN}{G_{\rm N}}
\newcommand{\Rm}{R_{\rm m}}
\newcommand{\bb}[1]{{\color{blue}#1}}

\begin{document}

\title{Supernova and neutron-star limits on 
large extra dimensions reexamined}

\author{Steen Hannestad}
\affiliation{Department of Physics, University of Southern Denmark,
Campusvej 55, 5230 Odense M, Denmark\\
and NORDITA, Blegdamsvej 17, 2100 Copenhagen, Denmark}

\author{Georg~G.~Raffelt} \affiliation{Max-Planck-Institut f\"ur
Physik (Werner-Heisenberg-Institut), F\"ohringer Ring 6, 80805
M\"unchen, Germany}

\date{3 April 2003, Erratum 13 Oct.\ 2003, {\color{blue}New Corrections (31 March 2025) in blue and see Appendix}}

\begin{abstract}
 In theories with large extra dimensions, supernova (SN) cores are
 powerful sources of Kaluza-Klein (KK) gravitons.  A large fraction of
 these massive particles are gravitationally retained by the newly
 born neutron star (NS). The subsequent slow KK decays produce
 potentially observable $\gamma$ rays and heat the NS.  We here show
 that the back-absorption of the gravitationally trapped KK gravitons
 does not significantly change our previous limits.  We calculate the
 graviton emission rate in a nuclear medium by combining the
 low-energy classical bremsstrahlung rate with detailed-balancing
 arguments. This approach reproduces the previous thermal emission
 rate, but it is much simpler and allows for a calculation of the
 absorption rate by a trivial phase-space transformation.  We derive
 systematically the dependence of the SN and NS limits on the number
 of extra dimensions.
\end{abstract}

\pacs{PACS numbers: 11.10.Kk, 98.70.Vc, 12.10.--g}

\maketitle

\section{Introduction}

Theories with large extra dimensions are a recent alternative to solve
the hierarchy problem of particle
physics~\cite{Arkani-Hamed:1998rs,Antoniadis:1998ig,%
Arkani-Hamed:1999nn,Han:1999sg,Giudice:1999ck,Hewett:cx}.  Within a
certain class of models, the most restrictive limits on the size of
the extra dimensions derive from the supernova (SN) emission of
Kaluza-Klein (KK) gravitons, particles with an essentially continuous
spectrum of masses that are a generic feature of the new theory.  Even
though these new particles interact very weakly, i.e.\ the strength of
ordinary gravitons, the number of modes and thus the size of the extra
dimensions is constrained by the requirement that SN~1987A did not
emit more KK gravitons than is compatible with the observed neutrino
signal duration~\cite{Cullen:1999hc%
,Barger:1999jf,Hanhart:2001er,Hanhart:2001fx}.

In the simplest models, KK~gravitons are stable except for their slow,
gravitational-strength decay into photons, neutrinos, and other
standard particles. Therefore, the decays of KK gravitons produced in
all cosmic SNe will contribute to the measured cosmic $\gamma$-ray
background, providing more restrictive limits than the SN~1987A
energy-loss argument~\cite{Hannestad:2001jv}.

Long after the parent SN has exploded, a neutron star (NS) will
continue to shine in $\gamma$-rays because a large fraction of the KK
gravitons remains gravitationally trapped---most of them were produced
with masses near the kinematical production threshold and thus with
small velocities.  Therefore, a NS is embedded in a halo of KK
gravitons that shines in 100~MeV $\gamma$-rays. The EGRET
non-observation of such signatures from nearby NSs thus provides
restrictive limits~\cite{Hannestad:2001xi}. The strongest constraints
yet, however, are from avoiding excess heating of certain old NSs
which otherwise could not cool to their observed low surface
temperatures~\cite{Hannestad:2001xi}.

These arguments depend on the assumption that KK gravitons are not
re-absorbed by inverse nuclear bremsstrahlung, a question that we
failed to address when we raised our argument~\cite{Hannestad:2001xi}.
A crude estimate of the re-absorption effect gives a time scale larger
than the ages of the stars that we used for our limits, but close
enough to warrant a more detailed calculation.  Moreover, if the
re-absorption time scale is shorter than the KK decay lifetime, NSs
will be heated directly by KK absorption rather than absorbing the
decay products. The excess heat would increase and our bounds would
improve.

The first goal of the present note is to calculate KK graviton
absorption in a NS. Essentially this is done by phase-space
transforming the emission rate.  This exercise is particularly simple
and transparent if one writes the graviton emission rate in a form
that separates the response of the thermal nuclear medium from the
phase space of the radiation. This approach significantly simplifies
the original calculation of the emission rate of Hanhart, Phillips,
Reddy and Savage~\cite{Hanhart:2001er} and illuminates the nature of
the approximations made. To achieve the same level of precision we
only need Weinberg's classical bremsstrahlung rate~\cite{Weinberg}
together with the principle of detailed balancing.  This approach
closely follows previous calculations of neutrino or axion emission
from a SN core based on the general properties of the thermal medium's
response functions~\cite{Raffelt:1993ix,Janka:1995ir}.  Therefore,
while the main goal of our derivation is to obtain the graviton
absorption rate, its derivation is an illuminating exercise in its own
right.

Our second goal is to extend our limits to the general case of $n$
extra dimensions.  Previously, the limits on the compactification
scale were explicitly stated only for the $n\leq3$ cases.  We use this
opportunity to show how the various SN and NS limits scale with $n$.

In Sec.~II we derive the emission and absorption of ordinary gravitons
in the low-energy limit from a nuclear medium.  In Sec.~III we extend
these results to the case of KK gravitons.  In Sec.~IV we revisit the
SN and NS constraints, and in Sec.~V we summarize our results.

\section{Nuclear Gravi-Bremsstrahlung}

\subsection{Differential Energy-Loss Rate}

The dominant graviton emission process from a SN core is
nucleon-nucleon bremsstrahlung $NN\to NNg$ where for now $g$ stands
for an ordinary graviton.  In the limit of soft radiation the
bremsstrahlung rate is determined by the measured nuclear scattering
cross section alone, i.e.\ the nuclear scattering event and the
associated bremsstrahlung process factorize~\cite{Hanhart:2001er}. In
the soft limit the details of the scattering process do not matter,
only the sudden change of the nucleons' energy-momentum tensor is
responsible for the radiation.

Assuming that two non-relativistic nucleons collide with the initial
CM momenta ${\bf p}_i={\bf p}_1=-{\bf p}_2$ to the final states ${\bf
p}_f={\bf p}_3=-{\bf p}_4$, the differential amount of emitted
gravi-bremsstrahlung energy is~\cite{Weinberg}
\begin{equation}
\frac{dE_g}{d\omega}=\frac{16\,G_{\rm N}}{5\pi}\,\,
\frac{{\bf p}_i^4+{\bf p}_f^4-2({\bf p}_i\cdot{\bf p}_f)^2}{M^2},
\end{equation}
where $G_{\rm N}$ is Newton's constant, $M$ the nucleon mass, and
$\omega$ the graviton energy.  We may also write this in terms of the
total initial and final kinetic nucleon energies, $E_i={\bf
p}_1^2/2M+{\bf p}_2^2/2M ={\bf p}_i^2/M$ and $E_f={\bf p}_3^2/2M+{\bf
p}_4^2/2M ={\bf p}_f^2/M$, and in terms of the CM scattering angle as
\begin{equation}\label{eq:dEgfull}
\frac{dE_g}{d\omega}=\frac{16\,G_{\rm N}}{5\pi}\,\,
\Bigl(E_i^2+E_f^2-2E_i E_f \cos^2\Theta_{\rm CM}\Bigr).
\end{equation}
Energy conservation implies $E_f=E_i-\omega$ so that
\begin{equation}\label{eq:dEgHanhart}
\frac{dE_g}{d\omega}=\frac{8\,G_{\rm N}}{5\pi}\,
(E_i+E_f)^2\sin^2\Theta_{\rm CM}
+{\cal O}(\omega^2)\,,
\end{equation}
an approximation used in Ref.~\cite{Hanhart:2001er}.

If the emission of soft gravitons ($\omega\to0$) is viewed as a
classical process, i.e.\ the colliding nucleons as external sources,
energy conservation does not apply, $E_f=E_i$, and we get the usual
flat bremsstrahlung spectrum,
\begin{equation}
\frac{dE_g}{d\omega}=\frac{32\,G_{\rm N}}{5\pi}\,
E_i^2\,\sin^2\Theta_{\rm CM}\,.
\end{equation}
Augmenting this classical result with detailed balancing will give us
the correct emission rate up to ${\cal O}(\omega)$ as in the more
complicated treatment of Ref.~\cite{Hanhart:2001er}.

Next, we introduce the nucleon scattering amplitude ${\cal A}$ which
is normalized such that the differential cross section for
neutron-neutron scattering is
\begin{equation}
\frac{d\sigma}{d\Omega}
=\frac{M^2\,\pi^4\,|{\cal A}|^2}{(2\pi)^6}\,.
\end{equation}
Note that $|{\cal A}|^2$ includes a spin sum over initial- and
final-state neutrons while $d\sigma/d\Omega$ is averaged over initial
and summed over final spins. If one takes $|{\cal A}|^2$ to be
independent of scattering angles, the total cross section~is
\begin{equation}
\sigma=\frac{M^2\,|{\cal A}|^2}{32\pi}=25~{\rm mb}\,.
\end{equation}
A factor $\frac{1}{2}$ for identical final-state nucleons was included
and the numerical value was quoted in Ref.~\cite{Hanhart:2001er} as a
good approximation for the conditions of interest.

With these ingredients the differential energy-loss rate of a neutron
medium in the soft limit is
\begin{equation}\label{eq:dQdw0}
\left.\frac{dQ}{d\omega}\right|_{\omega=0}
=\int d\Gamma S |{\cal A}|^2 \left[\frac{32\,G_{\rm N}}{5\pi}\,
E_i^2\,\sin^2\Theta_{\rm CM}\right]\,,
\end{equation}
where $S=\frac{1}{4}$ is a statistics factor for identical particles
in the initial and final state while $d\Gamma$ symbolizes the neutron
phase space integration, including all thermal occupation numbers,
blocking factors and the energy-momentum $\delta$ function.  True to
our soft-radiation approximation, the graviton energy-momentum does
not appear in this $\delta$ function.  Without the term in square
brackets Eq.~(\ref{eq:dQdw0}) is the rate of $nn$ collisions per unit
volume.

We now read off the dependence of the energy-loss rate $Q$ on various
input parameters. If the medium is non-degenerate, the $nn$ collision
rate per unit volume is the neutron number density squared times the
cross section times an average thermal relative velocity, the latter
being proportional to $(T/M)^{1/2}$ with $T$ the temperature. Further,
from the bremsstrahlung rate we have an average $E_i^2$ which is
proportional to $T^2$. A typical emitted graviton energy is of order
$T$ so that finally
\begin{equation}
Q\propto  G_{\rm N} \sigma\,n_B^2\,T^{7/2}\,M^{-1/2},
\end{equation}
where $n_B$ is the density of baryons (here neutrons).

We will perform explicit calculations only for the case of
non-relativistic non-degenerate neutrons. In that case the phase-space
integral can be transformed to one over CM
momenta~\cite{Hanhart:2001er}. Integrating a quantity $F$ over the
nucleon phase space yields
\begin{eqnarray}\label{eq:phasespace}
\int d\Gamma\,F&=&n_B^2\left(\frac{M}{\pi}\right)^{3/2}\,
\frac{T^{1/2}}{4}\nonumber\\
&\times&\int_0^\infty \!du_i\int_0^\infty \!du_f
(u_i u_f)^{1/2}\,e^{-u_i}\,\delta(u_i-u_f)
\nonumber\\
&\times&
\frac{1}{2}\int_{-1}^{+1}\!d\cos\Theta_{\rm CM} \,F\,,
\end{eqnarray}
where $u_{i,f}=E_{i,f}/T$. We have explicitly kept the energy $\delta$
function for later convenience.

Applying this result first to the $nn$ scattering rate per unit volume
we insert $F=S |{\cal A}|^2$. Assuming that the scattering amplitude
is independent of the CM energy and the scattering angle we find
explicitly
\begin{equation}
\Gamma_{nn}=\sigma\, n_B^2\, \left(\frac{4T}{\pi M}\right)^{1/2}
=\sigma\,\frac{n_B^2}{2}\,\langle v_{\rm rel}\rangle
\end{equation}
where the factor $\frac{1}{2}$ accounts for identical initial-state
particles. The corresponding final-state factor is already included
in the total cross section.

Next we perform the analogous integral for the differential
energy-loss rate and find
\begin{eqnarray}\label{eq:softemission}
\left.\frac{dQ}{d\omega}\right|_{\omega=0}
&=&\Gamma_{nn}\,\,\frac{128}{5\pi}\, G_{\rm N} T^2
\nonumber\\
&=&\frac{256}{5\pi^{3/2}}\,
\frac{G_{\rm N}\,\sigma\,n_B^2\,T^{5/2}}{M^{1/2}}
\end{eqnarray}
for the ``classical'' bremsstrahlung losses of the medium.

\subsection{Total Rate from Detailed Balancing}

The differential rate derived thus far is not yet useful for
calculating the total $Q$. To this end we observe that the emission,
absorption, or scattering of any radiation by any medium is described
by a suitable dynamical structure function. This is a quantum
correlator of those medium operators which couple to the relevant
radiation. For example, for neutrino pair emission, pair absorption,
or scattering, we need the nucleon spin correlator because neutrinos
couple primarily to the nucleon
spin~\cite{Raffelt:1993ix,Janka:1995ir}. For graviton emission we need
a dynamical structure function describing the microscopic fluctuations
of the medium's energy-momentum tensor.

However, in the soft limit when the energy transfer $\omega$ is small
we may use the long-wavelength limit where the radiation's momentum
transfer is ignored. The energy-loss rate of any type of radiation is
then of the form
\begin{equation}\label{eq:QS}
Q=\int \frac{d^3{\bf k}}{2\omega(2\pi)^3}\,\omega\,S(-\omega)
\end{equation}
where $S(\omega)$ is a function of the energy transfer alone.  Apart
from overall coefficients, $S$ is the dynamical structure function in
the long-wavelength limit.  The integral is over the invariant
phase-space of the radiation, one factor $\omega$ accounts for the
energy carried by the radiation.  Note that from the medium's
perspective a negative $\omega$ is energy lost, a positive $\omega$
energy gained. Therefore, bremsstrahlung emission involves
$S(-\omega)$.

The only property of relevance to our present discussion is the
detailed-balancing requirement
\begin{equation}
S(-\omega)=S(\omega)\,e^{-\omega/T}
\end{equation}
which is a general consequence of the non-commuting nature of the
correlated quantum operators describing the thermal medium, i.e.\ it
strictly is a quantum effect.  This condition ensures that the
efficiency of emission and absorption of radiation is such that in
equilibrium the radiation will reach a thermal distribution function.
As an immediate consequence the most general structure function is of
the form
\begin{equation}
S(\omega)=\bar S(\omega)\,\frac{2}{1+e^{-\omega/T}}
\end{equation}
where $\bar S(\omega)=\frac{1}{2}[S(\omega)+S(-\omega)]$ is symmetric
in $\omega$.

From Eq.~(\ref{eq:QS}) we may now write the differential energy-loss
rate in the form
\begin{equation}
\frac{dQ}{d\omega}=
\frac{\omega^2}{(2\pi)^2}\,\bar S(\omega)\,\frac{2}{1+e^{\omega/T}}.
\end{equation}
Comparing with Eq.~(\ref{eq:softemission}) we recognize that we may
write
\begin{equation}
\bar S(\omega)=\frac{S_0}{\omega^2}\,s(\omega/T)\,,
\end{equation}
where $s(x)$ is a dimensionless even function normalized to $s(0)=1$
and
\begin{eqnarray}
S_0&=&\frac{1024\,\pi^{1/2}}{5}\,
\frac{G_{\rm N}\,\sigma\,n_B^2\,T^{5/2}}{M^{1/2}}\nonumber\\
&=&1.547\times10^{16}~{\rm erg~cm^{-3}~s^{-1}~MeV^{-1}}
~T_{30}^{5/2}\,\rho_3^2\,.\nonumber\\
\end{eqnarray}
Here, $T_{30}=T/30~\rm MeV$ and $\rho_3=\rho/3\times10^{14}~\rm
g~cm^{-3}$.  Since $S(\omega)$ is even apart from the ``trivial''
detailed-balancing factor, we obtain the full differential energy-loss
rate up to first order in $x=\omega/T$ even though we have calculated
the bremsstrahlung rate only to zeroth order in $\omega/T$.

Our approach of using the concept of the structure function kills two
birds with one stone. We obtain the differential emission rate up to
order $\omega/T$ on the basis of the strictly classical soft-radiation
bremsstrahlung rate. Second, the inverse process, graviton absorption,
is simply a different phase-space integral over the same structure
function and thus trivial to extract.

A calculation up to ${\cal O}(\omega/T)$ is the best one can do in the
framework of the soft-radiation approximation. A more accurate
calculation would require modeling the nucleon interaction in detail
because the scattering and bremsstrahlung parts of the process no
longer factorize. Of course, a more precise calculation would also
require taking many-body effects into account, i.e.\ one would need to
calculate properly the dynamical structure function, an impossible
task at present.

In the absence of any more precise insights concerning the behavior
of the structure function we may use the simplest approximation
compatible with our level of approximation and take $s(x)=1$, dropping
all higher-order terms. In this case we find for the total emission
rate
\begin{equation}
Q=\frac{512\,\ln2}{5\pi^{3/2}}\,
G_{\rm N} \sigma\,n_B^2\,T^{7/2}\,M^{-1/2}\,.
\end{equation}
The coefficient is numerically 12.75.

The corresponding results of Ref.~\cite{Hanhart:2001er} are recovered
if in the nuclear phase-space integral Eq.~(\ref{eq:phasespace}) we
include the graviton energy, i.e.\ energy conservation now reads
$\delta(u_i-u_f-x)$ with $x=\omega/T$. Moreover, we must include the
expression Eq.~(\ref{eq:dEgHanhart}) for $dE_g/d\omega$. One first
performs $\int d\omega$ to remove the $\delta$ function and then
integrates the remaining expression. The coefficient of $Q$ is then
found to be $47104/(525\,\pi^{3/2})\approx16.11$, somewhat larger than
our result.

We can also extract the structure function implied by this
treatment. To this end we leave the $\int d\omega$ integration open
and rather do $\int du_i$ to remove the $\delta$ function. Collecting
all factors we find
\begin{eqnarray}
s(x)&=&\frac{1+e^{-x}}{48}
\int_0^\infty du_f(u_f^2+u_fx)^{1/2} e^{-u_f} (2u_f+x)^2\nonumber\\
&=&1+\frac{5}{48}\,x^2+{\cal O}(x^4).
\end{eqnarray}
However, $s(x)$ is not analytic, the coefficient of the $x^4$ term
diverges. This expression indeed agrees with ours up to ${\cal
O}(x)$. Put another way, up to ${\cal O}(\omega)$
Ref.~\cite{Hanhart:2001er} explicitly recovers the detailed-balancing
condition that we used as input information.

We can go through the same exercise for $dE_g/d\omega$ of
Eq.~(\ref{eq:dEgfull}) and find $s(x)=1+(5/16)\,x^2+{\cal
O}(x^4)$. This again deviates from the other results only at ${\cal
O}(x^2)$.

In summary, even though one has used the soft-radiation approximation
which is valid only for $\omega\ll T$, one obtains an estimate of the
emission rate which likely is correct within a numerical factor of
order~1. Still, this approach has essentially the status of an
``educated dimensional analysis.'' A real calculation would require
knowledge of the relevant dynamical structure function of the nuclear
medium, not just its long-wavelength, soft-radiation limit.

\subsection{Inverse Bremsstrahlung}

We are now in a position to calculate the graviton absorption rate by
a nuclear medium. For simplicity we may start with the energy-loss
rate Eq.~(\ref{eq:QS}), leave out the graviton phase space integration
except for the factor $1/2\omega$, remove the factor $\omega$ which
represents the emitted energy, include a factor 1/2 because the
energy-loss rate is summed over two graviton polarizations, and
substitute $\omega\to -\omega$ in the structure function because the
medium now gains energy. Then we find
\begin{eqnarray}\label{eq:absorptionrate}
\Gamma_g(\omega)&=&\frac{S(\omega)}{4\omega}\nonumber\\
&=&\frac{256\,\pi^{1/2}}{5}\,\,
\frac{G_N\sigma\,\rho^2\,T^{5/2}}{M^{5/2}}\,\,
\frac{2}{1+e^{-\omega/T}}\,\,\frac{1}{\omega^3}
\nonumber\\
&=&3.7\times10^{-17}~{\rm s}^{-1}
\rho_3^2\,\,T_{30}^{5/2}\,\omega_{100}^{-3}
\end{eqnarray}
where $\rho_3=\rho/3\times10^{14}~\rm g~cm^{-3}$, $T_{30}=T/30~\rm
MeV$, and $\omega_{100}=\omega/100~{\rm MeV}$.  We have assumed
$s(x)=1$ and in the numerical expression also $e^{-\omega/T}=0$.

\section{Kaluza-Klein Gravitons}

\subsection{Large Extra Dimensions}

Next we assume that space has $n$ large extra dimensions in which
gravity can propagate. String theory suggests a total of 11 space-time
dimensions so that there are at most 7 large extra dimensions in
addition to our 4-dimensional space-time. Therefore, we will limit our
explicit calculations and results to $1\leq n\leq 7$.

We further assume that the extra dimensions are toroidally
compactified with an equal radius $R$ for all of them. Put another
way, the linear size of each extra dimension is $2\pi R$ and the
volume of the compactified space is $V_n=(2\pi R)^n$.

Kaluza-Klein (KK) gravitons that propagate in the extra dimensions
with a momentum $p$ will appear to us as having a mass $m=p$.  The
modes are discrete with a density of states $V_n/(2\pi)^n= R^n$.  A
summation over all modes in the extra dimensions corresponds to
summing over all masses of KK gravitons in our world.  For large extra
dimensions the modes are narrowly spaced so that the sum over masses
transforms to an integral
\begin{equation}\label{eq:sumtointegral}
\sum_{\rm modes}\to \Omega_n R^n \int dm\,m^{n-1}\,,
\end{equation}
where
\begin{equation}
\Omega_n= \frac{2\pi^{n/2}}{\Gamma(n/2)}
\end{equation}
is the surface of the $n$-dimensional unit sphere.

Following the conventions of Ref.~\cite{Hewett:cx} we note that the
fundamental scale of the new theory is the effective Planck scale
$\overline M_{4+n}$ of the higher dimensional space.  It relates to
the conventional Planck scale $M_{\rm Pl}=G_{\rm
N}^{-1/2}=1.22\times10^{19}~\rm GeV$ by
\begin{equation}\label{eq:Mbardefine}
\overline M_{\rm Pl}^2= V_n \overline M_{4+n}^{2+n}
\end{equation}
where $\overline M_{\rm Pl}=M_{\rm Pl}/\sqrt{8\pi}$.  Numerically we
find
\begin{equation}
\frac{\overline M_{4+n}}{1~\rm TeV}=
\left(5.922\times10^{30}\right)^\frac{1}{2+n}
\left(\frac{3.140\times10^{-20}~\rm m}{R} \right)^\frac{n}{2+n}
\end{equation}
for the relationship between $\overline M_{4+n}$ and $R$.

Note that $\overline{M}_{4+n}$ is different from the energy scale
$M_{4+n}$ that was used by Hanhart {\it et al.}
\cite{Hanhart:2001er,Hanhart:2001fx} and that we called $M=M_{4+n}$ in
our previous papers~\cite{Hannestad:2001jv,Hannestad:2001xi}.  This
energy scale was defined from the relation
\begin{equation}\label{eq:Mdefine}
4 \pi G_{\rm N} = \left(\frac{1}{R M}\right)^n \frac{1}{M^2}\,.
\end{equation}
With Eq.~(\ref{eq:Mbardefine}) this implies
\begin{equation}
M = 2^{1/(n+2)} (2 \pi)^{n/(n+2)}\, \overline M_{4+n}\,.
\end{equation}
This relationship allows one to translate our new limits on $\overline
M_{4+n}$ into limits on the previous parameter $M$.  In the second row
of Table~\ref{tab:Mlimits} we give explicit values for $M/\overline
M_{4+n}$, i.e.\ one should multiply our new limits with this factor to
obtain the corresponding limits on the old parameter $M$.

\subsection{Absorption and Emission of KK Gravitons}

For KK gravitons, the absorption rate will be similar to that for
ordinary gravitons of the same energy.  Therefore, we may use
Eq.~(\ref{eq:absorptionrate}) directly for this purpose.

For the energy-loss rate we must take account of the large number of
available modes.  For a fixed KK mode with mass $m$ one has to include
in Eq.~(\ref{eq:QS}) a factor~\cite{Hanhart:2001er}
\begin{equation}
\left(\frac{19}{18}+\frac{11}{9}\,\frac{m^2}{\omega^2}
+\frac{2}{9}\,\frac{m^4}{\omega^4}\right)
\end{equation}
that already includes a summation over polarization states.  In
addition, $\int d^3{\bf k}\to 4\pi \int \omega\,k\,d\omega$ with
$k=(\omega^2-m^2)^{1/2}$.  The total emission rate then obtains by
summing over all masses.  With Eq.~(\ref{eq:sumtointegral}) one thus
finds
\begin{equation}\label{eq:Qn}
Q_{n}=\frac{\Omega_n R^n}{(2\pi)^2}
\int_0^\infty d\omega\,S(-\omega)\,\omega^{n+1}
\int_0^\omega dm\,G_{n-1}\left(\frac{m}{\omega}\right)\,,
\end{equation}
where we have defined
\begin{equation}
G_{p}(\mu)=
\mu^{p}\,\left(1-\mu^2\right)^{1/2}
\left(\frac{19}{18}+\frac{11}{9}\,\mu^2
+\frac{2}{9}\,\mu^4\right)\,,
\end{equation}
using $\mu=m/\omega$.
Performing the integral over masses explicitly yields
\begin{equation}\label{eq:Qnintegral}
Q_{n}=\frac{g_n R^n}{(2\pi)^2}
\int_0^\infty\!\! d\omega\,\omega^{n+2}\,S(-\omega)\,,
\end{equation}
where
\begin{equation}\label{eq:gndef}
g_{n}=\Omega_n\int_0^1 d\mu\,G_{n-1}(\mu)
=\frac{5\pi^{\frac{n+1}{2}}}{48}\,\frac{19+18n+3n^2}
{\Gamma(\frac{7+n}{2})}\,.
\end{equation}
Note that $g_0=19/18$ does not represent the emission of ordinary
gravitons because of differences in the summation over polarization
states.

With these results it is straightforward to state the differential
energy-loss rate. We read from Eq.~(\ref{eq:Qnintegral})
\begin{equation}
\frac{dQ_n}{d\omega}=\frac{S_0}{(2\pi)^2}\,
g_n(RT)^n\,\left(\frac{\omega}{T}\right)^n\,
\frac{2}{1+e^{\omega/T}}\,s(\omega/T)\,.
\end{equation}
Likewise
\begin{equation}
\frac{dQ_n}{d m}=\frac{S_0}{(2\pi)^2}\,
\Omega_n(RT)^n\,\left(\frac{m}{T}\right)^{n-1}\,F(m/T)\,,
\end{equation}
where
\begin{equation}
F(y)=\int_y^\infty dx\,\frac{2}{1+e^x}\,s(x)\,
G_0\left(\frac{y}{x}\right)
\end{equation}
with $y=m/T$ and $x=\omega/T$. For $s(x)=1$ a simple approximation is
$F(y)\approx2 y^{0.11}\,e^{-1.045\,y}$ that is good to about $\pm7\%$
for $0.1<\mu<20$.

If we finally use our expression for the structure function with
the simplest assumption $s(x)=1$ we find
\begin{equation}\label{eq:emissionrate}
Q_n=g_n f_n (RT)^n Q_0\,,
\end{equation}
where
\begin{eqnarray}
Q_0&=&\frac{512}{5\pi^{3/2}}\,
G_{\rm N} \sigma\,n_B^2\,T^{7/2}\,M^{-1/2}\nonumber\\
&=&2.35\times10^{16}\,{\rm erg~cm^{-3}~s^{-1}}
~T_{30}^{7/2}\,\rho_3^2\,.
\end{eqnarray}
Here, $T_{30}=T/30~\rm MeV$ and $\rho_3=\rho/3\times10^{14}~\rm
g~cm^{-3}$ and
\begin{equation}\label{eq:fndef}
f_n=(1-2^{-n})\Gamma(n+1)\,\zeta(n+1)\,,
\end{equation}
valid for $n\geq 1$.

In Table~\ref{tab:gnfn} we give numerical values for $(g_nf_n)^{1/n}$
because this is the effective $n$-dependent coefficient of $R$ in the
emission-rate formula. A simple approximation~is
\begin{equation}
(g_nf_n)^{1/n}\approx 1.81\,n^{0.425}\,,
\end{equation}
good to better than 1\% in the range $1\leq n\leq10$.

The corresponding emission rate of Ref.~\cite{Hanhart:2001er}, their
Eq.~(56), is systematically larger by factors of up to a few because
their $s(x)$ grows quadratically with $x$, increasing the emission of
higher-energy gravitons.  Recall that for the emission rate one needs
the structure function for $\omega/T$ of a few while the
soft-radiation approximation is only valid for $\omega\ll
T$. Therefore, our result agrees with Ref.~\cite{Hanhart:2001er} up to
the claimed accuracy of either result. The soft-radiation calculation
can not predict the emission rate to better than a factor of a few.

It is straightforward to calculate the average values of several
parameters of the emitted gravitons. In Table~\ref{tab:averages} we
give explicit numerical values for the average energy
$\langle\omega\rangle$, the average mass $\langle m\rangle$, and the
average velocity $\langle v\rangle$.  We also give the retention
fraction $f_{\rm trap}$ in a typical neutron star and the average mass
of the trapped gravitons as explained in the following Section.

\begin{table}[ht]
\caption{\label{tab:gnfn}Various coefficients described in the text.}
\begin{ruledtabular}
\begin{tabular}{lll}
$n$&$(g_n f_n)^{1/n}$&$(\Omega_n h_n)^{1/n}$\\
1&1.7943&0.3763\\
2&2.4544&0.9731\\
3&2.8981&1.1607\\
4&3.2597&1.2001\\
5&3.5771&1.1891\\
6&3.8662&1.1605\\
7&4.1348&1.1263\\
\end{tabular}
\end{ruledtabular}
\end{table}

\begin{table}[ht]
\caption{\label{tab:averages}Average properties of the emitted KK
gravitons.}
\begin{ruledtabular}
\begin{tabular}{llllll}
$n$&$\langle\omega\rangle/T$&$\langle m\rangle/T$&
$\langle v\rangle$&$f_{\rm trap}$&$\langle m\rangle_{\rm trap}/T$\\
1&1.1866 &0.5784&0.8061&0.0963&1.0902 \\
2&2.1923 &1.4054&0.7052&0.1816&2.0176 \\
3&3.1514 &2.2663&0.6384&0.2607&2.9052 \\
4&4.1060 &3.1516&0.5892&0.3342&3.7917 \\
5&5.0691 &4.0601&0.5506&0.4020&4.6890 \\
6&6.0429 &4.9888&0.5191&0.4643&5.5993 \\
7&7.0257 &5.9335&0.4927&0.5211&6.5208 \\
\end{tabular}
\end{ruledtabular}
\end{table}

\subsection{Neutron-Star Retention of KK Gravitons}

The KK gravitons are produced with barely relativistic velocities.
Therefore, the SN core will retain a large fraction of them
gravitationally after thermal production by the nuclear medium.  In
order to estimate the retention fraction we take the NS to be a
homogeneous sphere with a density $3\times10^{14}~\rm g~cm^{-3}$ and a
mass $M_{\rm NS}=1.4\,M_\odot$, corresponding to a radius of $R_{\rm
NS}=13.06$~km and a surface gravitational potential of
\begin{equation}
U_{\rm NS}=-\frac{G_{\rm N} M_{\rm NS}}{R_{\rm NS}}=-0.1139\,.
\end{equation}
We here use Newtonian physics rather than general relativity. Our
simple NS model leads to a radial dependence of the gravitational
potential within the NS of
\begin{equation}
U(r)=U_{\rm NS}\,\frac{3-r^2}{2}\,,
\end{equation}
where $r$ is a dimensionless radial coordinate that varies from 0 to 1
in the NS. A KK graviton produced at radius $r$ with energy $\omega$
remains trapped if its
kinetic plus gravitational energy is negative,
$U(r)+\frac{1}{2}v^2<0$ or
\begin{equation}
\omega\geq m\geq m_{\rm trap}(r)=\omega\,[1+U(r)]\,,
\end{equation}
where we have used that $|U|\ll 1$.

The graviton number emissivity is given by Eq.~(\ref{eq:Qnintegral})
if we drop one power of $\omega$ under the integral.
The number emissivity of those KK gravitons that remain trapped,
averaged over the entire NS, is found by
replacing $g_n$ with $g_n^{\rm trap}$ in these expressions.
The latter is found if in Eq.~(\ref{eq:gndef}) we substitute
\begin{equation}
\int_0^1 d\mu \,\ldots \to
3\int_0^1 dr\,r^2
\int_{\mu_{\rm trap}(r)}^1 d\mu\,\ldots\,,
\end{equation}
where $\mu_{\rm trap}=m_{\rm trap}/\omega=1+U(r)$.

For fixed $n$ the distribution of emitted masses is a universal
function of the parameter $\mu=m/\omega$, independently of the energy
$\omega$.  The velocity of a KK graviton is $v=(1-\mu^2)^{1/2}$ and
thus only a function of $\mu$ so that the distribution of KK
velocities is also independent of their energy. Therefore, the
trapping fraction is independent of the KK energy, in every energy
class $\omega$ the same fraction is retained given by $g_n^{\rm
trap}/g_n$. Explicit trapping fractions for different values of $n$
are shown in Table~\ref{tab:averages}.

As another consequence the average energy of the trapped KK gravitons
is identical with that of the escaping ones. However, the average mass
of those that remain trapped is much larger than the overall average.
We give explicit results for $\langle m\rangle_{\rm trap}$ in
Table~\ref{tab:averages}. These values are only slightly smaller than
$\langle \omega\rangle$, reflecting the fact that the trapped
particles are slow-moving and thus have only small amounts of kinetic
energy.

\section{Constraints on Large Extra Dimensions}

\subsection{Supernova 1987A}

We begin with the classical SN~1987A energy-loss limit on the emission
of KK gravitons~\cite{Cullen:1999hc%
,Barger:1999jf,Hanhart:2001er,Hanhart:2001fx}.  The duration of the
SN~1987A neutrino signal precludes that too much energy was emitted in
an invisible channel. For several cases involving axions or
right-handed neutrinos, self-consistent cooling calculations were
performed to determine the relationship between the allowed coupling
strength of the exotic particles and the duration of the neutrino
burst. The result can be summarized by the simple criterion that the
exotic energy-loss rate of a nuclear medium at a density of
$3\times10^{14}~\rm g~cm^{-3}$ and a temperature of 30~MeV should not
exceed $10^{19}~\rm
erg~g^{-1}~s^{-1}$~\cite{Raffelt:wa,Raffelt:1999tx}.  This criterion
is not crude or arbitrary, but is calibrated by numerical simulations
of different authors with different codes and for different cases and
reproduces these results surprisingly well.  The numerical studies of
Ref.~\cite{Hanhart:2001fx} for the specific case of KK gravitons
confirms once more the accuracy of the simple criterion.

We have only considered a non-degenerate neutron medium. For the
realistic case of a mixture of protons and neutrons the emission rate
would be somewhat larger, while degeneracy effects would slightly
decrease it. The soft-radiation calculation is only accurate to within
a factor of a few. Therefore, it is not worth worrying about the exact
influence of the mild nucleon degeneracy or the chemical
composition. The simple energy-loss criterion together with
Eq.~(\ref{eq:emissionrate}) thus implies
\begin{equation}
R\alt 6.58\times10^{-15}~{\rm m}~\left(1.28\times10^{17}\right)^{1/n}
(g_n f_n)^{-1/n}\,.
\end{equation}
We give the explicit limits for $1\leq n\leq 7$ in
Table~\ref{tab:Rlimits}. The corresponding limits on $\overline
M_{4+n}$ are given in Table~\ref{tab:Mlimits}.

For $n=2$ our limit $R\alt0.96~\mu\rm m$ compares with $0.70~\mu\rm m$
of Ref.~\cite{Hanhart:2001er} and $0.66~\mu\rm m$ of
Ref.~\cite{Hanhart:2001fx}.  For $n=3$ our limit $R\alt1.14~\rm nm$
compares with $0.83~\rm nm$ of Ref.~\cite{Hanhart:2001er} and $0.8~\rm
nm$ of Ref.~\cite{Hanhart:2001fx}.  The small differences are
perfectly in line with the different approximation made in the
calculation of the emission rate and with our using the simple
energy-loss criterion rather than a specific numerical simulation as
in Ref.~\cite{Hanhart:2001fx}.

\subsection{All Cosmic Supernovae}

In the toroidal compactification scheme assumed in our discussion the
only decay channels for KK gravitons are into those standard-model
particles that are kinematically allowed, notably into $2\gamma$,
$e^-e^+$ and $\nu\bar\nu$. The coupling is of gravitational strength
so that the mean lifetime is very large~\cite{Han:1998sg}
{\color{blue}
\begin{equation}\label{eq:lifetime}
\tau_{2\gamma}=\frac{1}{2}\tau_{e^-e^+}=
3.1\times10^{10}~{\rm yr}\,
\left(\frac{100~\rm MeV}{m}\right)^3\!.
\end{equation}}%
This is comparable to the age of the universe for masses in the
100~MeV range that would be typical for those KK gravitons emitted by
a SN core.

The photons from these decays from all past SNe would contribute to
the cosmic $\gamma$-ray background as measured by the EGRET
instrument~\cite{Kniffen1996}, constraining the amount of KK gravitons
that may have been emitted by all cosmic SNe~\cite{Hannestad:2001jv}.
The measured diffuse $\gamma$ background falls with energy roughly as
$E^{-2}$. We have explained in Ref.~\cite{Hannestad:2001jv} that this
particular power-law behavior implies that the limit on the KK
emission by SNe is nearly independent of the assumed core temperature
and nearly independent of the KK mass spectrum as long as the masses
are so large that most of the gravitons would have decayed within a
Hubble time. This is the case for $T\agt 20~\rm MeV$ and $n\agt2$.

In Ref.~\cite{Hannestad:2001jv} we found that the EGRET data imply
that a typical SN core must not lose more than about 0.5\% of its
energy in KK gravitons. With a realistic SN rate the limit would be
more restrictive by factors between 10 and 100. In the spirit of
deriving conservative limits it is thus justified to assume that for
all $n$ not more than 1\% of the energy loss is allowed that we used
for the SN~1987A limit. Put another way, the SN~1987A limits on $R$
scale with $(10^{-2})^{1/n}$. The corresponding explicit limits for
$1\leq n\leq 7$ are given in Tables~\ref{tab:Rlimits}
and~\ref{tab:Mlimits}, respectively.

\subsection{Supernova Remnant Cas~A}

The SN remnant Cas~A probably corresponds to Flamsteed's SN of 1680,
but in any case is so young that the cloud of emitted KK gravitons
would still appear as a point source to EGRET, even ignoring the fact
that a large fraction of them is gravitationally
retained~\cite{Hannestad:2001jv}.  The absence of an EGRET source at
the location of Cas~A implies that the flux of decay photons at Earth
from this source is limited by
\begin{equation}~\label{eq:pointsourcelimit}
\Phi_{E_\gamma>100~\rm MeV}\alt 10^{-7}~{\rm cm}^{-2}~{\rm s}^{-1}\,.
\end{equation}
In order to predict the photon flux at Earth we begin with
the KK emissivity Eq.~(\ref{eq:Qn}). To get the number
flux we must include a factor $\omega^{-1}$ under the integral.
Further, we must include the decay rate $\Gamma_{2\gamma}$,
involving the time-dilation factor $m/\omega$ and 
$\tau_{2\gamma}^{-1}$ from Eq.~(\ref{eq:lifetime}), i.e.\
\begin{equation}
\Gamma_{2\gamma}=\gamma\, \frac{m^4}{\omega}
\end{equation}
with {\color{blue}$\gamma=1.02\times10^{-24}~{\rm s^{-1}~MeV^{-3}}$}.
We further need
to multiply with the volume $V_{\rm NS}$ of the NS and with the
time-scale $\Delta t_{\rm NS}$ of KK emission, i.e.\ the cooling time
scale of the nascent neutron star.  We will use $V_{\rm NS}=M_{\rm
NS}/\rho$, assume a NS mass of $1.4\,M_\odot$, and take 
{\color{blue} $\Delta t_{\rm NS}=3~{\rm s}$.
This time scale corresponds to a realistic cooling time for the hot proto neutron star formed in a core collapse}. Finally we need to divide by $4\pi d^2$ with
$d=3.4~\rm kpc$ the distance to Cas~A.  We further need a factor 2 for
two decay photons per graviton.  We also assume that for every
decaying KK graviton we have $E_\gamma=\omega/2$, i.e.\ we ignore the
narrow energy distribution of the decay photons from non-relativistic
gravitons.
Collecting all factors we predict an EGRET flux at Earth of
\begin{eqnarray}\label{eq:CasAflux}
\Phi_{E_\gamma>E_0}&=&
\frac{\Omega_n R^n}{(2\pi)^2}\,
\frac{V_{\rm NS}\Delta t_{\rm NS}\gamma}{4\pi d^2}
\nonumber\\
&\times&2\int_{2E_0}^\infty d\omega\,S(-\omega)\, 
\int_0^\omega \!\!dm\, \omega^{n+3} G_{n+3}(m/\omega) \nonumber \\
\end{eqnarray}
where $E_0=100~\rm MeV$. The $\int dm$ integral is
explicitly $\omega^{n+4} h_n$ with
\begin{equation}
h_n=\frac{5n(278+223n+48n^2+3n^3)\sqrt{\pi}\Gamma(\frac{n}{2})}
{384\,\Gamma(\frac{n+11}{2})}\,.
\end{equation} 
Therefore, we have
\begin{equation}\label{eq:CasAflux2}
\Phi_{E_\gamma>E_0}=\Phi_0\,h_n\Omega_n (RT)^n\,\phi_n(2E_0/T)\,
\end{equation}
where
\begin{equation}
\phi_n(u)=\int_{u}^\infty dx\,\frac{x^{2+n}}{1+e^{x}}
\end{equation}
and
\begin{eqnarray}
\Phi_0&=&\frac{1}{(2\pi)^2}\,
\frac{V_{\rm NS}\Delta t_{\rm NS}\gamma 4S_0T^3}{4\pi d^2}
\nonumber\\
&=&
{\color{blue}
5.5\times10^{-25}~{\rm cm^{-2}~s^{-1}}}~
T_{30}^{11/2}\,\rho_3\,.
\end{eqnarray}
We given explicit values for $(\Omega_n h_n)^{1/n}$ in 
Table~\ref{tab:gnfn}.

Comparing with Eq.~(\ref{eq:pointsourcelimit}) then gives us the
limits on $R$ and $\overline M_{4+n}$ shown in
Tables~\ref{tab:Rlimits} and~\ref{tab:Mlimits}, respectively, for
$T=30$~MeV. In contrast to the previous results they now depend rather
sensitively on the SN core temperature.  Reducing $T$ to 20~MeV
degrades the $R$-limit for $n=2$ by about a factor of 10, for $n=7$
still by about a factor of~2.  The $\overline M_{4+n}$ limits are
somewhat less sensitive. Changing from 30 to 20~MeV degrades the $n=2$
limit by about a factor of 3.5, the $n=7$ limit by about a factor of
2. Overall the Cas~A limits are comparable to those from all past SNe.

\subsection{Gamma Radiation from Neutron Stars}

A large fraction of the KK gravitons emitted by a SN core are
gravitationally retained so that every NS would be embedded in a halo
of these particles. Therefore, NSs would be bright sources of
100~MeV $\gamma$-rays visible to EGRET.  The flux of decay photons
expected from a NS can be written in a way which is very similar
to Eq.~(\ref{eq:CasAflux}), except that the integral over $dm$ is
changed so that only the trapped gravitons are counted
\begin{equation}\label{eq:oldnsflux}
\Phi_{E_\gamma>E_0}=\Phi_0^* \, \Omega_n (RT)^n\,I_n(2E_0/T)\,,
\end{equation}
where 
\begin{equation}
\Phi_0^* =\bb{6.3\times10^{-24}~{\rm cm^{-2}~s^{-1}}}~
T_{30}^{11/2}\,\rho_3\, d_{\rm kpc}^{-2}\,.
\end{equation}
The function $I_n(u)$ is defined as
\bb{
\begin{eqnarray}
I_n(u) & = & \left(3\int_0^1 r^2 dr \int^1_{1+U(r)} d \mu\, G_{n+3}(\mu)\right)
\nonumber\\
&\times&\left(\int_u^\infty dx
\frac{x^{n+2}}{1+e^x}\right).
\end{eqnarray}}%
In Table~\ref{tab:I} we give $I_n$
for $n=1$--7 and $T=30$~MeV.

Comparing this flux prediction with the EGRET point-source limit
Eq.~(\ref{eq:pointsourcelimit}) for various old NSs then allows one to
derive limits on $R$.  In our previous paper~\cite{Hannestad:2001xi}
we used several old NSs for this argument.  The most restrictive
limits obtained from the nearest neutron star RX J185634-3754 that was
taken to be at a distance of 60~pc \cite{xray} and from PSR J0953+0755
at 120~pc \cite{tauris}.  However, the distance of RX J185634-3754 has
recently been adjusted upwards to $117 \pm 12$ pc \cite{RXnew}, so
that the distance is now exactly the same as for PSR J0953+0755. This
in turn means that the derived limits on $R$ from the gamma-ray
luminosity is exactly the same for the two stars.  The limits on $R$
and $\overline{M}_{4+n}$ thus derived are tabulated in Tables
\ref{tab:Rlimits} and \ref{tab:Mlimits}, respectively.

The flux prediction of Eq.~(\ref{eq:CasAflux2}) scales with the SN
core temperature as $T^{11/2}$, just as for the Cas~A argument.
Therefore, the limits are equally sensitive to $T$ as discussed in the
previous section.

\subsection{Neutron-Star Excess Heat}

\subsubsection{Decay of KK gravitons}

Even more stringent constraints obtain from considering the heating of
old NSs by the trapped cloud of KK gravitons surrounding
them~\cite{Hannestad:2001xi}.  When a graviton decays outside the NS,
the fraction of the decay photons hitting the surface of the star is
$\frac{1}{2}\sqrt{d^2-R_{\rm NS}^2}/d$, where $d$ is the distance of
the graviton from the center of the star, and $R_{\rm NS}$ is the NS
radius.

We proceed in the following way to calculate the rms distance of a
particle in the graviton cloud from the center of the star $\langle
d^2 \rangle^{1/2}$.  First we calculate the average total energy $E =
\frac{1}{2} v^2 + U(r)$ for the trapped gravitons. We then assume for
simplicity that all gravitons are on purely radial orbits. Given the
level of approximations entering this calculation this is certainly
justified.  By solving the equation of motion for a graviton with
energy $E$ on a radial orbit we get $\langle d^2 \rangle^{1/2}$, and
from this we can calculate the fraction, $\cal F$, of the total
KK-produced photon flux absorbed by the NS. In doing this we assume
that all gravitons decay when they are at a distance of exactly
$\langle d^2 \rangle^{1/2}$.  Numerical values for $\langle d^2
\rangle^{1/2}$ and $\cal F$ are shown in Table \ref{tab:nsheat1} for
$T=30$ MeV.

\begin{table}[t]
\caption{\label{tab:I}Values of the parameter $J_n$, and the
functions $\phi_n(2E_0/T)$ and
$I_n(2E_0/T)$, taken for $T = 30$~MeV and $E_0=100$~MeV.}
\begin{ruledtabular}
\begin{tabular}{llll}
$n$&$\phi_n(2E_0/T)$ & $I_n(2E_0/T)$ & $J_n$ \\
1& 0.605 & {\color{blue}0.0458} & {\color{blue}1.765} \\
2& 4.933 & {\color{blue}0.345}  & {\color{blue}8.28} \\
3& 41.42 & {\color{blue}2.69}   & {\color{blue}46.3} \\
4& 360.2 & {\color{blue}21.7}   & {\color{blue}302.2} \\
5& $3.27 \times 10^3$ &{\color{blue}182.7} & {\color{blue}$2.25\times10^3$}\\
6& $3.11 \times 10^4$ & {\color{blue}$1.62\times10^3$} & {\color{blue}$1.89\times10^4$}\\
7& $3.13 \times 10^5$ & {\color{blue}$1.52\times10^4$} & {\color{blue}$1.76\times10^5$} \\
\end{tabular}
\end{ruledtabular}
\end{table}

\begin{table}[ht]
\caption{\label{tab:nsheat1}Values of $\langle d^2 \rangle^{1/2}$,
$\cal F$, $t_{\rm inside}$, and the reabsorption rate 
$\Gamma_n$.}
\begin{ruledtabular}
\begin{tabular}{lllll}
$n$&$\langle d^2 \rangle^{1/2}$&$\cal F$ & 
$t_{\rm inside}$&$\Gamma_n~[{\rm s}^{-1}]$ \\
1& 1.405 & 0.149 & 0.269 & $1.22 \times 10^{-16}$ \\
2& 1.372 & 0.158 & 0.279 & $2.06 \times 10^{-17}$ \\
3& 1.352 & 0.164 & 0.287 & $7.18 \times 10^{-18}$ \\
4& 1.320 & 0.174 & 0.297 & $3.44 \times 10^{-18}$ \\
5& 1.291 & 0.184 & 0.308 & $1.94 \times 10^{-18}$ \\
6& 1.262 & 0.195 & 0.319 & $1.21 \times 10^{-18}$ \\
7& 1.236 & 0.206 & 0.330 & $8.14 \times 10^{-19}$ \\
\end{tabular}
\end{ruledtabular}
\end{table}

Having calculated the fraction of the total flux absorbed by
the star, we can calculate the total energy per unit time
absorbed by the star, $L$, as
\begin{eqnarray}
\label{eq:nsheat}
L&=&\frac{\Omega_n R^n}{(2\pi)^2}\, V_{\rm NS}\Delta t_{\rm NS}\gamma
{\cal F} \,\, \bb{3} \int_0^1 r^2 dr
\nonumber \\
&& \, \times \int_{0}^\infty d\omega\,  \omega^{n+5} 
\, S(-\omega)\, 
\bb{\int^1_{1+U(r)} \!\!d\mu \, G_{n+3}(\mu)} \nonumber \\
& = &\bb{4.53 \times 10^{-18} }L_\odot \,\, \Omega_n (RT)^n \, {\cal F} 
\, J_n \, 
\rho_3 \, T_{30}^{13/2}.
\end{eqnarray}
Values for the parameter
\bb{
\begin{eqnarray}
J_n& = & \left(3\int_0^1 r^2 dr \int^1_{1+U(r)} d \mu\, G_{n+3}(\mu)\right)
\nonumber\\
&\times&\left(\int_0^\infty dx
\frac{x^{n+3}}{1+e^x}\right)
\end{eqnarray}}%
are shown Table~\ref{tab:I}.

\begin{table*}
\caption{\label{tab:Rlimits}Upper limits on the compactification
radius $R$ (in meters) from our arguments.}
\medskip
\begin{ruledtabular}
\begin{tabular}{llllllll}
$n$&1&2&3&4&5&6&7\\
Neutrino Signal\\
\quad SN~1987A&
$4.9\times10^{2}$&$1.0\times10^{-6}$&$1.1\times10^{-9}$&
$3.8\times10^{-11}$&$4.9\times10^{-12}$&$1.2\times10^{-12}$&
$4.4\times10^{-13}$\\
EGRET $\gamma$-ray limits\\
\quad All cosmic SNe&
4.9&$1.0\times10^{-7}$&$2.5\times10^{-10}$&$1.2\times10^{-11}$&
$1.9\times10^{-12}$&$5.6\times10^{-13}$&$2.3\times10^{-13}$\\
\quad Cas A&
\bb{$5.3\times10^{3}$}&\bb{$1.3\times10^{-6}$}&\bb{$9.2\times10^{-10}$}&
\bb{$2.6\times10^{-11}$}&\bb{$3.1\times10^{-12}$}&\bb{$7.6\times10^{-13}$}&
\bb{$2.8\times10^{-13}$}\\
\quad RX J185635--3754&
\bb{$16.$}&\bb{$6.7\times10^{-8}$}&\bb{$1.2\times10^{-10}$}&\bb{$5.6\times10^{-12}$}&
\bb{$9.0\times10^{-13}$}&\bb{$2.7\times10^{-13}$}&\bb{$1.1\times10^{-13}$}\\
Neutron-star excess heat\\
\quad PSR J0953+0755&
\bb{$8.3$}&\bb{$5.9 \times 10^{-8}$}&\bb{$1.3 \times 10^{-10}$}&
\bb{$5.9\times10^{-12}$}&\bb{$9.4 \times 10^{-13}$} &\bb{$2.8 \times 10^{-13}$}&
\bb{$1.2 \times 10^{-13}$} \\
\end{tabular}
\end{ruledtabular}
\end{table*}

\begin{table*}
\caption{\label{tab:Mlimits}Lower limits on the fundamental energy
scale $\overline M_{4+n}$ (in TeV) corresponding to the $R$-limits of
Table~\ref{tab:Rlimits}.  Multiply limits on $\overline M_{4+n}$ by
the factor $M/\overline{M}_{n+4}$ given in the second row to obtain
limits on the parameter $M$ used in our previous papers
\protect\cite{Hannestad:2001jv,Hannestad:2001xi}.}
\medskip
\begin{ruledtabular}
\begin{tabular}{llllllll}
$n$&1&2&3&4&5&6&7\\
$M/\overline{M}_{n+4}$ & 2.32 & 2.98 & 3.46 & 3.82 & 4.10 & 4.32 
& 4.51 \\
Neutrino Signal\\
\quad SN~1987A&
$7.4\times10^2$&8.9&0.66&$0.12$&$3.5\times10^{-2}$
&$1.44\times10^{-2}$&$7.2\times10^{-3}$\\
EGRET $\gamma$-ray limits\\
\quad All cosmic SNe
&$3.4\times10^{3}$&28.&1.7&$0.25$&$6.8\times10^{-2}$
&$2.6\times10^{-2}$&$1.2\times10^{-2}$\\
\quad Cas A&
\bb{$3.3\times10^{2}$}&\bb{7.7}&\bb{0.74}&\bb{$0.15$}&\bb{$4.8\times10^{-2}$}
&\bb{$2.0\times10^{-2}$}&\bb{$1.0\times10^{-2}$}\\
\quad RX J185635--3754&
\bb{$2.2 \times 10^3$}&\bb{34.}&\bb{2.5}&\bb{0.42}&\bb{0.12}&\bb{$4.5\times 10^{-2}$}
&\bb{$2.1 \times 10^{-2}$}\\
Neutron-star excess heat\\
\quad PSR J0953+0755&
\bb{$2.8 \times 10^3$} &\bb{36.} & \bb{2.5} & \bb{0.41} & \bb{0.11} &\bb{$4.3 \times 10^{-2}$} &
\bb{$2.0\times 10^{-2}$} \\
\end{tabular}
\end{ruledtabular}
\bigskip
\end{table*}

The thermal cooling time scale of an isolated neutron star is of order
$10^5$--$10^6$~yr. The neutron star PSR J0953+0755 has an estimated
age of $17 \times 10^6$~yr, and HST observations indicate that it has
a total luminosity of roughly $10^{-5} L_\odot$ \cite{ll99}.
Comparing this to the total energy transferred to the star from KK
graviton decay, Eq.~(\ref{eq:nsheat}), yields an upper limit on $R$.
In Tables \ref{tab:Rlimits} and \ref{tab:Mlimits} we have tabulated
limits on $R$ and $\overline{M}_{4+n}$ derived from the maximum energy
transfer per unit time to the neutron star $L_{\rm max} = 10^{-5}
L_\odot$. As we discussed in Ref.~\cite{Hannestad:2001xi}, this limit
is far stronger than other astrophysical limits on $R$. For $n < 5$ it
is stronger than any current laboratory limits.

The predicted NS heating rate of Eq.~(\ref{eq:nsheat}) scales with the
SN core temperature as $T^{13/2}$, slightly steeper than for the
previous arguments.  The $T$-sensitivity of the limits on $R$ and
$\overline M_{4+n}$ is almost the same as before.

\subsubsection{Reabsorption in the neutron star}

KK gravitons in the trapped cloud will be inside the NS on part of
their trajectory. Assuming, as before, that all gravitons move on
radial orbits we have calculated the fraction of time an average
graviton spends inside the NS, $t_{\rm inside}$.  While the graviton
is inside the neutron star it can potentially be reabsorbed by the
nuclear medium. Using Eq.~(\ref{eq:absorptionrate}) we find the
following absorption rates for average gravitons
\begin{equation}
\Gamma_{\rm reabsorption} = \Gamma_n  \,
\rho_3^2 \, T_{30}^{-1/2}\,,
\end{equation}
where values of $\Gamma_n$ are tabulated in Table~\ref{tab:nsheat1}.
Comparing this reabsorption rate to the age of the two NSs RX
J185634-3754 ($1.6 \times 10^{13}$ s) and PSR J0953+0755 ($5.4 \times
10^{14}$ s) we see that reabsorption in the medium is at most a
moderate effect.

\newpage

\section{Conclusions}

We have systematically revisited the constraints on KK graviton
emission by SN cores and neutron stars that have been discussed in the
literature. We have paid close attention to the scaling with the
number $n$ of extra dimensions and to the conventions concerning the
relationship between the radius $R$ of the large extra dimensions and
the fundamental energy scale, the $4+n$ dimensional effective Planck
mass $\overline M_{4+n}$. Our limits are summarized in
Tables~\ref{tab:Rlimits} and~\ref{tab:Mlimits}. For $n=2$ and 3 the
limits based on EGRET observations
are slightly less restrictive than stated in our previous papers,
largely because we use a somewhat smaller emission rate from a SN
core, and because the estimated distance to RX
J185635-3754 has increased by a factor of~2.
However, the bound from neutron star excess heat is in fact slightly
more restrictive than stated in our previous paper. The reason is that
the average distance of a graviton from the neutron star center is
somewhat lower in the present calculation, compared to what was used
in our previous work.

The main theoretical new ingredient of our work is a new approach to
calculating the emission rate. For low-energy gravitational
bremsstrahlung the emission rate can be calculated directly from the
classical formula in conjunction with the universal detailed-balancing
property of the medium's dynamical structure function. This
calculation agrees with previous works to the specified order of
precision, but is vastly simpler and clearly illuminates the nature of
the used approximation.  The KK graviton emission rate by a SN core
can not be calculated to better than a factor of a few. However, this
uncertainty hardly affects the limits, especially for large $n$,
because the limiting value of $R$ is proportional to the $n$-th root
of the limiting emission rate.

Another benefit of our derivation is that it produces directly the
absorption rate. It is found to be so small that the gravitationally
trapped KK gravitons are not reabsorbed too quickly to invalidate our
previous arguments.


\begin{acknowledgments}
In Munich, this work was partly supported by the Deut\-sche
For\-schungs\-ge\-mein\-schaft under grant SFB 375 and the ESF network
Neutrino Astrophysics. G.R.\ thanks the Scuola Internazionale
Superiore di Studi Avanzati (SISSA), Trieste, for hospitality during a
visit where this work was begun. S.H.\ thanks the MPI f{\"u}r Physik
for hospitality while this paper was finished.
\end{acknowledgments}

\newpage

\onecolumngrid

\section*{CORRECTIONS 31 MARCH 2025}

\twocolumngrid

Almost a quarter century after this paper was written, interest in these questions has returned \cite{Lust:2019zwm, Montero:2022prj, Anchordoqui:2022txe, Anchordoqui:2025nmb}. In this context, Krzysztof Jod{\l}owski and Jamie Law-Smith have independently pointed out an unfortunate inconsistency. The constraints from neutron star excess heat (the last rows in Tables~V and VI) do not follow from Eq.~(55) if one compares this predicted luminosity with the nominal bound of $10^{-5}\,L_\odot$. The stated limits in the Tables are reproduced, within rounding errors, if the right-hand side of Eq.~(55) is multiplied with a factor of 276. However, Eq.~(55) numerically agrees with its derivation and is not misprinted. We cannot explain how this factor might have crept in when evaluating Eq.~(55).

However, looking at the old paper with greater scrutiny turns up several further issues, consisting of some outright errors by us and previous authors, some thoughts on additional effects that should be considered, and a rough estimate of what credible bounds might be. 

Corrections of outright mistakes or direct updates of parameters are fixed in the above text and the modifications are marked in blue, while in the following we explain what the modifications are.

\subsection{KK decay rate}

The numerical lifetime of KK gravitons stated in Eq.~(44) was taken from Eq.~(47) of Ref.~\cite{Han:1998sg}, where the coefficient was stated as $6\times10^9~{\rm y}$, which is what we used. This result contains several mistakes. In their post-publication v4 of the arXiv posting, the decay rate was increased by a factor of 2, so their corrected decay rate for KK gravitons $h\to\gamma\gamma$ is
\begin{eqnarray}\label{eq:KK-photon-decay}
    \Gamma_{h\to\gamma\gamma}&=&\frac{\GN m_h^3}{10}
    \nonumber\\
    &=&1.02\times10^{-18}~{\rm s}^{-1}\,\left(\frac{m_h}{100~{\rm MeV}}\right)^3
    \nonumber\\
    &=&\frac{1}{3.1\times 10^{10}~{\rm y}}\,\left(\frac{m_h}{100~{\rm MeV}}\right)^3.
\end{eqnarray}
However, in all versions of their paper, the numerical evaluation is wrong by a factor of 10, the lifetime in v4 being stated as $3\times10^9~{\rm y}$. In other words, after correcting the factor of 2, the numerical lifetime is longer by a factor of 5 compared to what was originally stated, leading to our modified Eq.~(44). We observe that in their Eq.~(58) for a different channel, the numerical evaluation is correct and in direct contradiction with the numerical evaluation leading to their Eq.~(47).

The decay rate for $h\to e^+e^-$ stated in their Eq.~(55) has also changed and is again twice as fast than the photon decay, i.e., the lifetime $\tau_{e^+e^-}=2\tau_{\gamma\gamma}$ as stated in our Eq.~(44). On the other hand, the decay rate into neutrinos, that we previously also stated, is actually not provided in Ref.~\cite{Han:1998sg}. For Dirac neutrinos, it should be the same as for electrons. For Majorana neutrinos, there are fewer states and possible helicity constraints, so this case is less obvious.

The rough bounds from the cosmic diffuse $\gamma$ rays from all past SNe would not change much since the main assumption was that all KK gravitons would have decayed over the age of the universe, which is still approximately true, considering that there are also other decay channels. Therefore, we have not modified the stated bounds in Tables~V and VI coming from all cosmic SNe.

The numerical decay rate reduced by a factor of 5 then also leads to a corresponding reduction of the parameter $\gamma$ stated after our Eq.~(46), which is numerically the decay rate of a 1~MeV KK graviton.

\subsection{SN cooling time}

The restrictive constraints on KK emission mean that their overall impact on SN physics is a small perturbation. All of our arguments are based on a simple one-zone model of the SN core, assuming homogeneous nuclear density and fixed $T$ for the entire emission period, that we took as $\Delta t_{\rm NS}=7.5~{\rm s}$, inspired by the SN~1987A signal duration. However, the cooling of a SN core (or proto neutron star) is probably faster due to the effect of convective energy transport \cite{Fiorillo:2023frv}. Therefore, a duration for KK production of 
\begin{equation}\label{eq:coolingtime}
    \Delta t_{\rm NS}=3~{\rm s}
\end{equation}
is probably more realistic, reducing the number of produced KK gravitons roughly by a factor of~2.

\subsection{Flux limit from Cas~A}

In this way, the numerical coefficient in Eq.~(55) for the $\gamma$-ray flux is reduced by a bit more than a factor of 10. The limits stated in Tables~V and VI get reduced correspondingly. For small $n$, these modified limits are worse than those from the SN~1987A signal and in this sense not self-consistent---the energy loss would here be dominated by KK gravitons and not by neutrinos, i.e., the KK emission would not be a small perturbation. In any event, the Cas A bound is not competitive and we have corrected it only for numerical consistency.

\subsection{Gamma radiation from neutron stars}

These modification of the $\gamma$-ray flux also affects the prediction from old neutron stars. The baseline flux in Eq.~(53) is the same as Eq.~(51) except for the reference distance of now 1~kpc so that Eq.~(53) is Eq.~(51) times $(3.4\,{\rm kpc}/1\,{\rm kpc})^2$, providing the revised value of $\Phi_0^*$.

However, there are additional corrections. The function $I_n(u)$ in Eq.~(54) was originally defined as
\begin{equation}
I_n(u)=\frac{1}{3} \int_0^1 r^2 dr \int_u^\infty dx
\frac{x^{n+2}}{1+e^x} \int_0^{1-U(r)} d \mu\, G_{n+3}(\mu).
\end{equation}
However, the average over emission radius should instead be
$3\int_0^1r^2 dr$ as in Eq.~(42) so that, if nothing under the integral depends on $r$, it would be unity. Moreover, the integral over masses $d\mu$ must have $1+U(r)$ on one end, not $1-U(r)$, because the gravitational potential $U(r)$ is negative and the quantity $\mu=m/\omega$ cannot exceed unity. Indeed, the original tabulated values are reproduced with this typographical sign correction. However, the integral is over the escaping gravitons, not the trapped ones, although both populations are a fraction of order unity. Notice that for $\mu=m/\omega=1$, gravitons are at rest and do not escape, whereas those with $\mu=0$ have infinite energy. So finally the integral should be from $1+U(r)$ to 1 as written in the corrected Eq.~(54). We write it with a different typographical grouping because the energy integral does not depend on $r$. The new $I_n$ values are shown in blue in Table~III. They are a factor of 6--21 larger than the original values, which includes a factor of 9 from turning $1/3$ to~3.

Another issue concerns the neutron-star distances. We stated a common distance of 120~pc for the two neutron stars PSR J0953+0755 and RX J185635–3754, but mistakenly had used 60~pc, a correction that was the subject of our first Erratum and v2 on arXiv. However, for PSR J0953+0755, also known as B0950+08, the older distance estimate of 120~pc was superseded by a VLBA measurement of $162\pm5$~pc \cite{Brisken:2002ri}, reducing the flux by almost a factor of~2, and so we strike this pulsar from the list. The distance of RX J185635–3754 has since been confirmed and is now stated as $123^{+11}_{-15}$~pc \cite{Walter:2010ht}, so we keep the original 120~pc for this case.

\subsection{Neutron star excess heat}

Similar issues arise for the definition of $J_n$ in Eq.~(56), where likewise the original factor $1/3$ should be 3 and the limits of integration likewise modified. The new $J_n$ values are shown in blue in Table~III. Once more, they are a factor of 6--21 larger than the original values. Moreover, the numerical coefficient in Eq.~(55) also changes due to the modified decay rate and cooling time.

From a modern perspective, the luminosity limit of $10^{-5}\,L_\odot$ for the pulsar PSR J0953+0755 (aka B0950+08) appears far too restrictive to derive a limit. (In Tables V and VI, its name was misprinted as PSR J0952+0755.) Recent determinations of the surface temperature give a range $T=1$--$3\times10^5$~K \cite{Pavlov:2017eeu}. Assuming a radius of 13~km, corresponding to our simple model, implies a blackbody luminosity range of 3--$300\times10^{-5}\,L_\odot$. The lower end is not very different from our earlier value, but the upper end, more appropriate for a limit, is much larger.

The corrected limits based on $300\times10^{-5}\,L_\odot$ are shown in Tables~V and VI. They are comparable to those from the other arguments and no longer much more restrictive. 

On the other hand, this pulsar is too warm relative to a normal cooling history. It is not excluded that KK decays actually heat it to some degree and that this could be the explanation for its excess heat.

\subsection{Sanity check}

Because the neutron-star heating bound has strongly changed, a quick sanity check is warranted, using back-of-the-envelope estimates instead of detailed calculations. To this end, we concentrate on $n=1$, which apparently is of main contemporary interest.

The total energy emitted by the SN core in the form of KK gravitons is
\begin{equation}
    E_{\rm KK}=g_nf_n (RT)^n Q_0 \frac{M_{\rm NS}}{\rho_{\rm nuc}}\Delta t_{\rm NS},
\end{equation}
where the energy-loss rate per unit volume was given in Eqs.~(35) and (36), $M_{\rm NS}=1.4\,M_\odot$, $\rho_{\rm nuc}=3\times10^{14}~{\rm g}~{\rm cm}^{-3}$, and $\Delta t_{\rm NS}=3$~s following  Eq.~(\ref{eq:coolingtime}). For $n=1$, we find $f_n g_n=25\pi^3/432=1.794$ as in Table~I. Since we always use $T=30$~MeV, we may write numerically $(R T)^n=(\Rm\,1.521{\times}10^{14})^n$, where $\Rm$ is $R$ in meters. Numerically, for $n=1$, we thus find
\begin{equation}
    E_{\rm KK}=\Rm\, 1.80\times10^{50}~{\rm erg}.
\end{equation}
As a scale for a typical emitted KK mass we take $m=T=30$~MeV, which is exactly what is found for the average trapped mass given in the first line of Table~II. The decay time of Eq.~(44) is then $\tau_{2\gamma}=3.6\times10^{19}$~s and the $\gamma$-ray luminosity of the total emitted KK energy is
\begin{equation}
    L_{\rm KK}=\Rm\, 5.0\times10^{30}~{\rm erg}~{\rm s}^{-1}=
    \Rm\,1.2\times10^{-3}\,L_\odot.
\end{equation}
The trapping fraction and the amount of this luminosity hitting the NS are factors of order unity, let's say 10\% of this luminosity heats the NS. Comparing with an excess heat of $10^{-3}\,L_\odot$ yields a constraint of $\Rm\alt 10$, in agreement with the corrected bound in Table~V. The corrections are: A decay rate faster by a factor of 10. An assumed heat excess a factor of 100 smaller. The unexplained factor of 276 mentioned at the beginning of this Erratum. Therefore, the old bound was more restrictive by the factor $10\times100\times276$, almost exactly corresponding to what was stated in the original Table~V.

\onecolumngrid

\section*{GALACTIC BULGE (CORRECTIONS FOR REF.~\texorpdfstring{\cite{Casse:2003pj}}{})}

\twocolumngrid

Shortly after our paper had appeared, other authors pointed out that the $\gamma$ rays produced by KK decays from all pulsars in the galactic bulge would provide the most restrictive limits \cite{Casse:2003pj}. Moreover, these would not depend on individual objects, avoiding possible peculiarities of any specific case and its observations. The galactic center is at a distance of 8~kpc, compared to RX J185635–3754 at 0.12~kpc, so one loses a flux factor of $(0.12\,{\rm kpc}/8\,{\rm kpc})^2=2.3\times10^{-4}$, but the large number of 4--$10\times10^8$ pulsars strongly overcompensates this loss. The smaller number 4, appropriate for a limit, implies gaining a factor $0.9\times10^5$ relative to RX J185635–3754.

However, the stellar population in the GB is old as stressed in Ref.~\cite{Casse:2003pj} and star formation was much more active at early times. Therefore, most of the GB neutron stars are old and many of the KK gravitons trapped around them would have decayed before the present epoch. Decays forming at least 100~MeV $\gamma$-rays require KK masses of at least 200~MeV. The relevant decay channels are those into $\gamma\gamma$, $e^+e^-$, $\mu^+\mu^-$, and $\nu\bar\nu$ of all flavors. For yet larger masses, pion channels also become important. Without pions and assuming Dirac neutrinos, the decay rate is $(1+5/2)=3.5$ times the $2\gamma$ one, i.e., the lifetime is $1/3.5$ the one given in Eq.~(44).  In this case, the decay time is $3.1\times10^{10}~{\rm yr}/3.5/8=1.1\times10^9~{\rm yr}$. Assuming an age of 13.8 billion years, the exponential suppression is $\exp(-13.8/1.1)=4\times10^{-6}$. For more massive KK states, the decay would be even faster. Therefore, the oldest NSs are completely irrelevant.

Therefore, the result depends sensitively on the time profile of NS formation. If one were to assume a constant birthrate of $\dot N_{\rm NS}=4\times10^8/(13.8\,{\rm Gyr})$, a lifetime of $\tau=1.1$~Gyr implies a survival of \smash{$\dot N_{\rm NS}\tau=3\times10^7$}, meaning that only 10\% of the originally assumed pulsars are today available. However, the star formation rate was larger in the past and what counts are those pulsars formed during the past Gyr. If the present-day birthrate were only 10\% of the average would reduce the relevant pulsars by another factor of ten. Therefore, the original gain factor of about $10^6$ would be reduced to $f_{\rm gain}=10^4$, still a formidable improvement over using one close pulsar.

The BG bounds on $R$ would be those of the RX J185635–3754 line in Table~V, improved by $f_{\rm gain}^{-n}$ and thus by far the most restrictive. However, what exactly is $f_{\rm gain}$ depends on the number of NSs in the GB formed over the past Gyr, a number that is not readily available.

However, for $n=1$, there is an additional reduction of sensitivity due to reabsorption of KK gravitons. The reabsorption rate in Table~IV of 0.26~Gyr means a further reduction of the number of relevant pulsars by another factor of~4, and the reabsorption rate may be yet larger by the pionic processes discussed below.

Still, the GB pulsars could still provide the most restrictive limit and deserve a dedicated reexamination.

\onecolumngrid
\section*{POSSIBLE FUTURE IMPROVEMENTS}

\twocolumngrid

If there is renewed interest in bounds of this type, one obvious direction is to revisit all of these arguments from a modern perspective and contemporary input parameters. The theoretical calculations of the KK decay rates of Ref.~\cite{Han:1998sg} should be independently checked. Moreover, among additional low-energy channels of relevance here is the pionic one that affects KK gravitons with masses exceeding $2m_\pi=280$~MeV.

The Fermi Gamma-Ray Space Telescope (formerly GLAST) has been launched since our paper was written and has taken data for a long time, superseding the EGRET observations. The $\gamma$-ray flux limits for some nearby pulsars should be much more restrictive than those from EGRET, perhaps by a few orders of magnitude. A quick glance at the literature reveals, however, that while $\gamma$-ray observations of many pulsars have been published, flux limits on those which are not observed in $\gamma$-rays are not readily available. Naturally, the precise non-observation of a given object, of main interest here, is not important to $\gamma$-ray astronomers. Therefore, one would have to dig more deeply into this question and perform an analysis probably along the lines of Ref.~\cite{Berenji:2016jji}, but today with a much longer exposure.

In addition, $NN$ bremsstrahlung need not be the dominant emission process. In the context of axion emission, the pionic process $\pi^- +p\to n+a$ was found to emit a few times more energy than bremsstrahlung, and with a much harder spectrum \cite{Carenza:2020cis}. For graviton emission, one would need to calculate the rate for this process, because it could lead to significant improvements. The harder spectrum would allow for the production of more massive KK gravitons and faster decay.

Concerning reabsorption of the gravitationally trapped KK states, the main concern of this paper, the pionic process $h+n\to p+\pi^-$ or $h+n\to n+\pi^0$ may dominate instead of inverse bremsstrahlung. For $n=1$, the absorption time from Table~IV was found to be 260~Myr, much larger than the age of the used pulsars. If the true absorption rate is a factor of a few larger, and the absorption time correspondingly smaller, is probably still large enough to make reabsorption a small concern, but this question should be quantitatively checked. In principle, reabsorption would be another contribution to the heating of pulsars besides KK decay, a question that may deserve more detailed attention.

For $n=1$, the limits on $R$ given in Table~V are 5--10 meters, so bringing them down to the sub-mm or even micron scale would require an improvement exceeding four orders of magnitude that seems unlikely. Still, a fresh analysis with modern inputs would be called for.

\subsection*{Acknowledgments}

We thank Krzysztof Jod{\l}owski and Jamie Law-Smith for calling our attention to the inconsistency mentioned at the beginning of this Erratum, G\"unter Sigl for helpful comments on an earlier version of our comments on the Galactic Bulge issue, and Dieter L\"ust for stressing the present-day renewed relevance of astrophysical limits on KK gravitons.



\begin{thebibliography}{00}

\bibitem{Arkani-Hamed:1998rs}
N.~Arkani-Hamed, S.~Dimopoulos and G.~Dvali,
``The hierarchy problem and new dimensions at a millimeter,''
Phys.\ Lett.\ B {\bf 429}, 263 (1998)
[\href{https://arxiv.org/abs/hep-ph/9803315}{hep-ph/9803315}].

\bibitem{Antoniadis:1998ig}
I.~Antoniadis, N.~Arkani-Hamed, S.~Dimopoulos and G. Dvali,
``New dimensions at a millimeter to a Fermi and superstrings at a TeV,''
Phys.\ Lett.\ B {\bf 436}, 257 (1998)
[\href{https://arxiv.org/abs/hep-ph/9804398}{hep-ph/9804398}].

\bibitem{Arkani-Hamed:1999nn}
N.~Arkani-Hamed, S.~Dimopoulos and G.~Dvali,
``Phenomenology, astrophysics and cosmology of theories with  
sub-millimeter dimensions and TeV scale quantum gravity,''
Phys.\ Rev.\ D {\bf 59}, 086004 (1999)
[\href{https://arxiv.org/abs/hep-ph/9807344}{hep-ph/9807344}].

\bibitem{Han:1999sg}
T.~Han, J.~D.~Lykken and R.~Zhang,
``On Kaluza-Klein states from large extra dimensions,
Phys.\ Rev.\ D {\bf 59}, 105006 (1999)
[\href{https://arxiv.org/abs/hep-ph/9811350}{hep-ph/9811350}].

\bibitem{Giudice:1999ck}
G.~F.~Giudice, R.~Rattazzi and J.~D.~Wells,
``Quantum gravity and extra dimensions at high-energy colliders,''
Nucl.\ Phys.\ B {\bf 544}, 3 (1999)
[\href{https://arxiv.org/abs/hep-ph/9811291}{hep-ph/9811291}].

\bibitem{Hewett:cx}
J.~Hewett and J.~March-Russell,
``Extra Dimensions,''
in: K.~Hagiwara et al.\ (Particle Data Group),
``The Review of Particle Physics,''
Phys.\ Rev.\ D {\bf 66}, 010001 (2002).

\bibitem{Cullen:1999hc}
S.~Cullen and M.~Perelstein,
``SN1987A constraints on large compact dimensions,''
Phys.\ Rev.\ Lett.\ {\bf 83}, 268 (1999)
[\href{https://arxiv.org/abs/hep-ph/9903422}{hep-ph/9903422}].

\bibitem{Barger:1999jf}
V.~Barger, T.~Han, C.~Kao and R.~J.~Zhang,
``Astrophysical constraints on large extra dimensions,''
Phys.\ Lett.\ B {\bf 461}, 34 (1999)
[\href{https://arxiv.org/abs/hep-ph/9905474}{hep-ph/9905474}].

\bibitem{Hanhart:2001er}
C.~Hanhart, D.~R.~Phillips, S.~Reddy and M.~J.~Savage,
``Extra dimensions, SN~1987A, and nucleon nucleon scattering data,''
Nucl.\ Phys.\ B {\bf 595}, 335 (2001)
[\href{https://arxiv.org/abs/nucl-th/0007016}{nucl-th/0007016}].

\bibitem{Hanhart:2001fx}
C.~Hanhart, J.~A.~Pons, D.~R.~Phillips and S.~Reddy,
``The likelihood of GODs' existence: Improving the SN~1987A 
constraint on  the size of large compact dimensions,''
Phys.\ Lett.\ B {\bf 509}, 1 (2001)
[\href{https://arxiv.org/abs/astro-ph/0102063}{astro-ph/0102063}].

\bibitem{Hannestad:2001jv}
S.~Hannestad and G.~Raffelt,
``New supernova limit on large extra dimensions:
Bounds on Kaluza-Klein graviton production,''
Phys.\ Rev.\ Lett.\ {\bf 87}, 051301 (2001)
[\href{https://arxiv.org/abs/hep-ph/0103201}{hep-ph/0103201}].

\bibitem{Hannestad:2001xi}
S.~Hannestad and G.~G.~Raffelt,
``Stringent neutron-star limits on large extra dimensions,''
Phys.\ Rev.\ Lett.\  {\bf 88}, 071301 (2002)
[\href{https://arxiv.org/abs/hep-ph/0110067}{hep-ph/0110067}].

\bibitem{Weinberg}
S.~Weinberg,
``Gravitation and Cosmology''
(Wiley, New York, 1972).

\bibitem{Raffelt:1993ix}
G.~Raffelt and D.~Seckel,
``A selfconsistent approach to neutral current processes 
in supernova cores,''
Phys.\ Rev.\ D {\bf 52}, 1780 (1995)
[\href{https://arxiv.org/abs/astro-ph/9312019}{astro-ph/9312019}].

\bibitem{Janka:1995ir}
H.-T.~Janka, W.~Keil, G.~Raffelt and D.~Seckel,
``Nucleon spin fluctuations and the supernova emission 
of neutrinos and axions,''
Phys.\ Rev.\ Lett.\ {\bf 76}, 2621 (1996)
[\href{https://arxiv.org/abs/astro-ph/9507023}{astro-ph/9507023}].

\bibitem{Raffelt:wa}
G.~G.~Raffelt,
``Stars As Laboratories For Fundamental Physics,''
(University of Chicago Press, 1996).

\bibitem{Raffelt:1999tx}
G.~G.~Raffelt,
``Particle physics from stars,''
Ann.\ Rev.\ Nucl.\ Part.\ Sci.\  {\bf 49}, 163 (1999)
[\href{https://arxiv.org/abs/hep-ph/9903472}{hep-ph/9903472}].

\bibitem{Han:1998sg}
T.~Han, J.~D.~Lykken and R.~J.~Zhang,
``On Kaluza-Klein states from large extra dimensions,''
Phys.\ Rev.\ D {\bf 59}, 105006 (1999)
[\href{https://arxiv.org/abs/astro-ph/hep-ph/9811350}{hep-ph/9811350}].

\bibitem{Kniffen1996}
D.~A.~Kniffen et al.,
``EGRET Observations of the high latitude diffuse radiation,''
Astron. Astrophys. Suppl. Ser. {\bf 120}, 615 (1996).

\bibitem{xray}
F.~M.~Walter,
``The proper motion, parallax, and origin of the isolated 
neutron star RX J185635-3754,''
Astrophys.\ J.\ {\bf 549}, 433 (2001).
[\href{https://arxiv.org/abs/astro-ph/0009031}{astro-ph/0009031}].

\bibitem{tauris}
T.~M.~Tauris {\it et al.},
``Discovery of PSR J0108-1431: The closest known neutron star?,''
Astrophys.\ J.\ Lett.\ {\bf 428}, L53 (1994).

\bibitem{RXnew}F.~M.~Walter and J.~M.~Lattimer,
``A revised parallax and its implications for RX J185635-3754,''
Astrophys.\ J.\ Lett.\ {\bf 576}, L145 (2002).

\bibitem{ll99}
M.~B.~Larson and B.~Link,
``Superfluid friction and late-time thermal evolution of 
neutron stars,''
Astrophys.\ J.\ {\bf 521}, 271 (1999)
[\href{https://arxiv.org/abs/astro-ph/9810441}{astro-ph/981044}].

\bibitem{Lust:2019zwm}
D.~L{\"u}st, E.~Palti and C.~Vafa,
``AdS and the Swampland,''
Phys. Lett. B \textbf{797}, 134867 (2019)
[\href{https://arxiv.org/abs/1906.05225}{1906.05225}].

\bibitem{Montero:2022prj}
M.~Montero, C.~Vafa and I.~Valenzuela,
``The dark dimension and the Swampland,''
JHEP \textbf{02}, 022 (2023)
[\href{https://arxiv.org/abs/2205.12293}{2205.12293}].

\bibitem{Anchordoqui:2022txe}
L.~A.~Anchordoqui, I.~Antoniadis and D.~L{\"u}st,
``Dark dimension, the swampland, and the dark matter fraction composed of primordial black holes,''
Phys. Rev. D \textbf{106}, 086001 (2022)
[\href{https://arxiv.org/abs/2206.07071}{2206.07071}].

\bibitem{Anchordoqui:2025nmb}
L.~Anchordoqui, I.~Antoniadis and D.~L{\"u}st,
``Two Micron-Size Dark Dimensions,''
[\href{https://arxiv.org/abs/2501.11690}{2501.11690}].

\bibitem{Fiorillo:2023frv}
D.~F.~G.~Fiorillo, M.~Heinlein, H.-T.~Janka, G.~Raffelt, E.~Vitagliano and R.~Bollig,
``Supernova simulations confront SN 1987A neutrinos,''
Phys. Rev. D \textbf{108}, 8 (2023)
[\href{https://arxiv.org/abs/2308.01403}{2308.01403}].

\bibitem{Brisken:2002ri}
W.~F.~Brisken, J.~M.~Benson, W.~M.~Goss and S.~E.\ Thorsett,
``Very Long Baseline Array Measurement of Nine Pulsar Parallaxes,''
Astrophys. J. \textbf{571}, 906 (2002)
[\href{https://arxiv.org/abs/astro-ph/0204105}{astro-ph/0204105}].

\bibitem{Walter:2010ht}
F.~M.~Walter, T.~Eisenbeiss, J.~M.~Lattimer, B.~Kim, V.~Hambaryan and R.~Neuhaeuser,
``Revisiting the Parallax of the Isolated Neutron Star RX J185635-3754 Using HST/ACS Imaging,''
Astrophys. J. \textbf{724}, 669 (2010)
[\href{https://arxiv.org/abs/1008.1709}{1008.1709}].

\bibitem{Pavlov:2017eeu}
G.~G.~Pavlov, B.~Rangelov, O.~Kargaltsev, A.~Reisenegger, S.~Guillot and C.~Reyes,
``Old but still warm: Far-UV detection of PSR B0950+08,''
Astrophys. J. \textbf{850}, 79 (2017)
[\href{https://arxiv.org/abs/1710.06448}{1710.06448}].

\bibitem{Casse:2003pj}
M.~Cass{\'e}, J.~Paul, G.~Bertone and G.~Sigl,
``Gamma-rays from the galactic bulge and large extra dimensions,''
Phys. Rev. Lett. \textbf{92}, 111102 (2004)
[\href{https://arxiv.org/abs/hep-ph/0309173}{hep-ph/0309173}].

\bibitem{Carenza:2020cis}
P.~Carenza, B.~Fore, M.~Giannotti, A.~Mirizzi and S.~Reddy,
``Enhanced Supernova Axion Emission and its Implications,''
Phys. Rev. Lett. \textbf{126}, 071102 (2021)
[\href{https://arxiv.org/abs/2010.02943}{2010.02943}].

\bibitem{Berenji:2016jji}
B.~Berenji, J.~Gaskins and M.~Meyer,
``Constraints on Axions and Axionlike Particles from Fermi Large Area Telescope Observations of Neutron Stars,''
Phys. Rev. D \textbf{93}, 045019 (2016)
[\href{https://arxiv.org/abs/1602.00091}{1602.00091}].

\end{thebibliography}
\end{document}